\documentclass[
 reprint,
 superscriptaddress,
 amsmath,amssymb,
 aps,
 prl
]{revtex4-2}

\usepackage{gensymb}
\usepackage{graphicx}
\usepackage{dcolumn}
\usepackage{bm}
\usepackage{hyperref}
\usepackage[mathlines]{lineno}
\usepackage{verbatim}
\usepackage{siunitx}
\usepackage[italic]{hepnames}

\begin{document}

\null\hfill\begin{tabular}[t]{l@{}}
\small{FERMILAB-PUB-21-296-ND-T}
\end{tabular}

\title{New constraints on tau-coupled Heavy Neutral Leptons with masses $m_N = 280-970$\,MeV}

\author{R.~Acciarri}
\affiliation{Fermi National Accelerator Lab, Batavia, Illinois 60510, USA}

\author{C.~Adams}
\affiliation{Argonne National Lab, Lemont, Illinois 60439, USA}

\author{J.~Asaadi}
\affiliation{University of Texas at Arlington, Arlington, Texas 76019, USA}

\author{B.~Baller}
\affiliation{Fermi National Accelerator Lab, Batavia, Illinois 60510, USA}

\author{V.~Basque}
\affiliation{Fermi National Accelerator Lab, Batavia, Illinois 60510, USA}

\author{F.~Cavanna}
\affiliation{Fermi National Accelerator Lab, Batavia, Illinois 60510, USA}

\author{A.~{de Gouv\^ea}}
\affiliation{Northwestern University, Evanston, Illinois 60208, USA}

\author{R.S.~Fitzpatrick}
\affiliation{University of Michigan, Ann Arbor, Michigan 48109, USA}

\author{B.~Fleming}
\affiliation{Yale University, New Haven, Connecticut 06520, USA}

\author{P.~Green}
\email{patrick.green-3@postgrad.manchester.ac.uk}
\affiliation{University of Manchester, Manchester M13 9PL, United Kingdom}

\author{C.~James}
\affiliation{Fermi National Accelerator Lab, Batavia, Illinois 60510, USA}

\author{K.J.~Kelly}
\affiliation{Fermi National Accelerator Lab, Batavia, Illinois 60510, USA}

\author{I.~Lepetic}
\affiliation{Rutgers University, Piscataway New Jersey 08854, USA}

\author{X.~Luo}
\affiliation{University of California, Santa Barbara, California, 93106, USA}

\author{O.~Palamara}
\affiliation{Fermi National Accelerator Lab, Batavia, Illinois 60510, USA}

\author{G.~Scanavini}
\affiliation{Yale University, New Haven, Connecticut 06520, USA}

\author{M.~Soderberg}
\affiliation{Syracuse University, Syracuse, New York 13244, USA}

\author{J.~Spitz}
\affiliation{University of Michigan, Ann Arbor, Michigan 48109, USA}

\author{A.M.~Szelc}
\affiliation{University of Edinburgh, Edinburgh EH9 3FD, United Kingdom}

\author{W.~Wu}
\affiliation{Fermi National Accelerator Lab, Batavia, Illinois 60510, USA}

\author{T.~Yang}
\affiliation{Fermi National Accelerator Lab, Batavia, Illinois 60510, USA}

\collaboration{The ArgoNeuT Collaboration}
\noaffiliation

\begin{abstract}
A search for Heavy Neutral Leptons has been performed with the ArgoNeuT detector exposed to the NuMI neutrino beam at Fermilab. We search for the decay signature $N \to \nu \mu^+ \mu^-$, considering decays occurring both inside ArgoNeuT and in the upstream cavern. In the data, corresponding to an exposure to $\num{1.25e20}$ POT, zero passing events are observed consistent with the expected background. This measurement leads to a new constraint at 90\% confidence level on the mixing angle $\left\vert U_{\tau N}\right\rvert^2$ of tau-coupled Dirac Heavy Neutral Leptons with masses $m_N =$ 280 - 970\,MeV, assuming $\left\vert U_{eN}\right\rvert^2 = \left\vert U_{\mu N}\right\rvert^2 = 0$.
\end{abstract}

\maketitle

\emph{Introduction}.--- The discovery that neutrinos oscillate and therefore have mass has inspired numerous experimental efforts to understand this phenomenon. The Standard Model (SM) does not predict the existence of neutrino masses, requiring additional fields and/or interactions to generate them. One such model requires the existence of two or more Heavy Neutral Leptons (HNLs): SM gauge singlet fermions that mix with the light neutrinos. This mixing can induce the observed small neutrino masses via one of many different seesaw mechanisms~\cite{Minkowski:1977sc,GellMann:1980vs,Yanagida:1979as,Glashow:1979nm,Mohapatra:1979ia,Schechter:1980gr,Foot:1988aq}. In addition, HNLs can provide solutions to other mysteries of nature such as the baryon asymmetry of the universe~\cite{Davidson:2008bu} (via leptogenesis) or dark matter~\cite{Boyarsky:2018tvu}. In this paper, we present a search for HNLs with masses $\mathcal{O}(100)$\,MeV using the ArgoNeuT detector.

We consider the simplest phenomenological scenario including a HNL, $N$ -- that it has a mass $m_N$ and mixes with the light neutrinos via one or more non-zero new angles $\left\vert U_{eN}\right\rvert^2$, $\left\vert U_{\mu N}\right\rvert^2$, and $\left\vert U_{\tau N}\right\rvert^2$ in an extended $4\times 4$ leptonic mixing matrix. If $m_N$ is in the ${\sim}$MeV-GeV range, HNLs can be produced as a result of high-energy proton--fixed-target collisions, travel to a downstream detector and decay producing detectable charged particles. In the ArgoNeuT detector, we search for the decay signature $N \to \nu \mu^+ \mu^-$. 

ArgoNeuT was a 0.24 ton Liquid Argon Time Projection Chamber (LArTPC) neutrino detector located in the NuMI beam \cite{Adamson:2015dkw} at Fermilab that collected data in 2009-2010. The instrumented volume of the TPC was $40 \times 47 \times 90$\,cm$^3$ (vertical, drift, beam direction) with two readout planes, each consisting of 240 wires spaced by $4$\,mm and oriented at $\pm60 \degree$ to the horizontal. A detailed description of the design and operation of the ArgoNeuT detector can be found in Ref.~\cite{Anderson:2012vc}. The ArgoNeuT detector was located $100$\,m underground in the MINOS near detector hall, $1033$\,m downstream of the NuMI target and immediately upstream of the MINOS near detector (MINOS-ND). ArgoNeuT was able to use the MINOS-ND as a muon spectrometer. A detailed description of the MINOS-ND can be found in Ref.~\cite{Michael:2008bc}. The analysis reported in this paper is performed using \num{1.25e20} protons-on-target (POT) collected in reverse horn current (anti-neutrino) mode, during which both ArgoNeuT and the MINOS-ND were operational \cite{Anderson:2012vc}.

\emph{Generation and simulation}.--- 
A HNL with mass $m_N$ and mixing angle $\left\lvert U_{\alpha N}\right\rvert^2$ with the light neutrinos, $\nu_\alpha$ ($\alpha = e,\mu,\tau$), can be produced by any kinematically accessible process that would normally result in an outgoing $\nu_\alpha$. For the decay $N \to \nu \mu^+ \mu^-$ we require $m_N > 2m_\mu$ and, for simplicity, we assume that only one angle $\left\lvert U_{\alpha N}\right\rvert^2$ is non-zero at a time. A variety of experiments~\cite{Bergsma:1985is,Bernardi:1985ny,Bernardi:1987ek,Vaitaitis:1999wq,Artamonov:2009sz,Abe:2019kgx,Abratenko:2019kez,NA62:2020mcv,CortinaGil:2021gga} have set powerful constraints on the angles $\left\lvert U_{eN}\right\rvert^2$ and $\left\lvert U_{\mu N}\right\rvert^2$ in the region of interest for ArgoNeuT. We therefore focus on the case where $\left\lvert U_{\tau N}\right\rvert^2$ is the only non-zero mixing angle. In this scenario, the HNLs are predominantly produced in the decays of $\tau^\pm$ leptons originating from decays of $D_{(s)}^\pm$ mesons. The lifetime of $N$, as well as the branching ratio $\mathrm{Br}(N\to \nu\mu^+ \mu^-)$, can be calculated as a function of the mixing $\left\vert U_{\tau N}\right\rvert^2$ considering all kinematically accessible final states~\cite{Gorbunov:2007ak}. We assume that $N$ is a Dirac fermion throughout this analysis.

In the NuMI beam approximately 87\% of the incident $120$\,GeV protons interact in the target, with the majority of the remaining 13\% interacting $715$\,m downstream in the hadron absorber~\cite{Adamson:2015dkw}. We consider HNL production occurring in both the target and the absorber, the latter giving access to shorter $N$ lifetimes as a result of being significantly closer to the detector. We simulate the particle propagation using GEANT4~\cite{Agostinelli:2002hh} and the $\tau^\pm$ production using PYTHIA8~\cite{Sjostrand:2014zea}. Approximately 10\% of the beam protons reach the absorber with energy $T_p \approx 120$\,GeV. For $120$\,GeV protons interacting in either the target or the absorber, an average of $2.1 \times 10^{-7}$ ($3.0 \times 10^{-7}$) $\tau^+$ ($\tau^-$) are produced per proton~\cite{Coloma:2020lgy}. To generate a flux of $N$, we simulate the decays $\tau^\pm \to N X$, where $X$ consists of SM particles. We simulate the kinematics by assuming $m_X = m_{\pi^\pm}$ and that the branching ratio of this new decay is $\mathrm{Br}(\tau^\pm \to N X^\pm) = 0.9 \left\lvert U_{\tau N}\right\vert^2 K(m_N)$~\cite{Gorbunov:2007ak, footnote:1}. Since the $D_{(s)}^\pm$ and $\tau^\pm$ lifetimes are small, the kinematics of $N$ produced in the target and absorber are qualitatively the same. However, the geometric acceptance of ArgoNeuT is significantly larger for the absorber-produced $N$ due to the proximity to the detector.

The HNL decay products are then simulated in the ArgoNeuT detector using the LArSoft software framework \cite{Snider:2017wjd}, which simulates the particle propagation using GEANT4 \cite{Agostinelli:2002hh} then performs detector response simulation and reconstruction \cite{Anderson:2012vc,Acciarri:2018ahy}. A stand-alone version of the MINOS simulation and reconstruction is then used to simulate the tracks exiting ArgoNeuT and entering the MINOS-ND.  

\begin{figure}%[h]
  \centering
  \includegraphics[width=.45\textwidth,keepaspectratio]{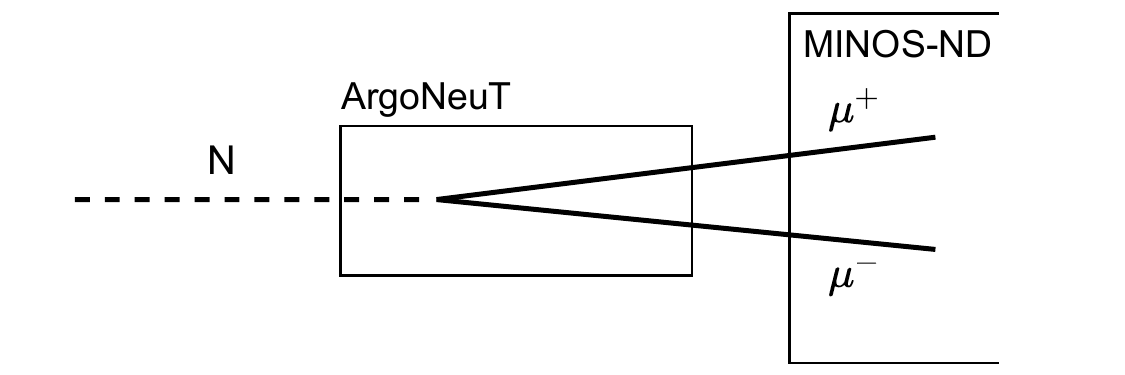}
  \includegraphics[width=.45\textwidth,keepaspectratio]{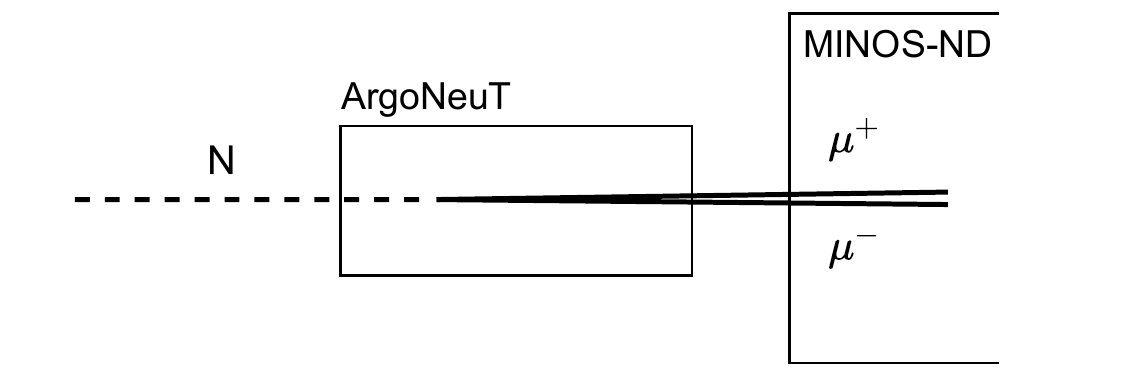}
  \includegraphics[width=.45\textwidth,keepaspectratio]{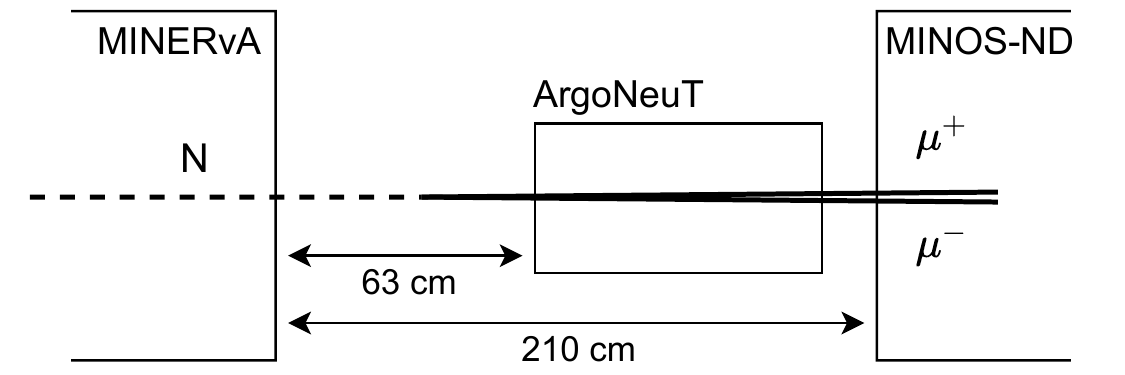}
  \caption{Diagrams of HNL decays occurring inside the ArgoNeuT detector with the two-track (top) and double-MIP (middle) decay signatures, and of a decay occurring in the cavern between ArgoNeuT and MINERvA with the double-MIP signature (bottom). Diagrams not to scale.}
  \label{fig:signature_diagram}
\end{figure}

\emph{Signature}.--- 
The HNL decay $N \rightarrow \Pnu \APmuon \Pmuon$ is seen in ArgoNeuT as a pair of minimally ionising particles (MIPs) that can be matched to a pair of oppositely charged particles in the MINOS-ND. These muons are energetic and highly forward-going: with average energy $\langle E_{\mu^\pm} \rangle \sim 7 $\,GeV; average angle with respect to the beam direction $\langle \theta_{beam} \rangle \sim 1.5 \degree$; and an average opening angle $\langle \theta_{opening} \rangle \sim 3 \degree$. Given the ArgoNeuT angular resolution of approximately $3 \degree$ \cite{spitzthesis}, this results in the muon pair frequently overlapping and being reconstructed as a single track for part or all of their length. Two in-ArgoNeuT decay signatures are therefore considered, each of which is illustrated in Fig. \ref{fig:signature_diagram} (top, middle). In the first, the muons are reconstructed as two distinct MIP tracks originating from a common vertex, each of which can be matched to tracks in the MINOS-ND. This signature will be referred to as a {\it two-track} event. In the second, the muons overlap and are reconstructed as a single track with double-MIP $dE/dx$ for part or all of their length. Then, in the MINOS-ND, the pair of oppositely-charged muons separate due to the presence of a magnetic field. This signature will be referred to as a {\it double-MIP} event.

In addition to decays occurring inside the ArgoNeuT detector, we also consider decays occurring in the cavern upstream of ArgoNeuT along the NuMI beam-line where the resulting muons then pass through the detector. This scenario is illustrated in Fig. \ref{fig:signature_diagram} (bottom). During the ArgoNeuT physics run, the MINERvA detector \cite{Aliaga:2013uqz} was under construction in the upstream cavern. We therefore only consider decays that occur in the 63\,cm between the end of the MINERvA detector and the start of the ArgoNeuT TPC. In this scenario, only the double-MIP signature is considered. This is because the two-track signature is more difficult to distinguish from neutrino-induced background muons due to the absence of vertex information, whereas the double-MIP signature is unique to potential HNL decays.

\begin{figure}%[h]
  \centering
  \includegraphics[width=.5\textwidth,keepaspectratio]{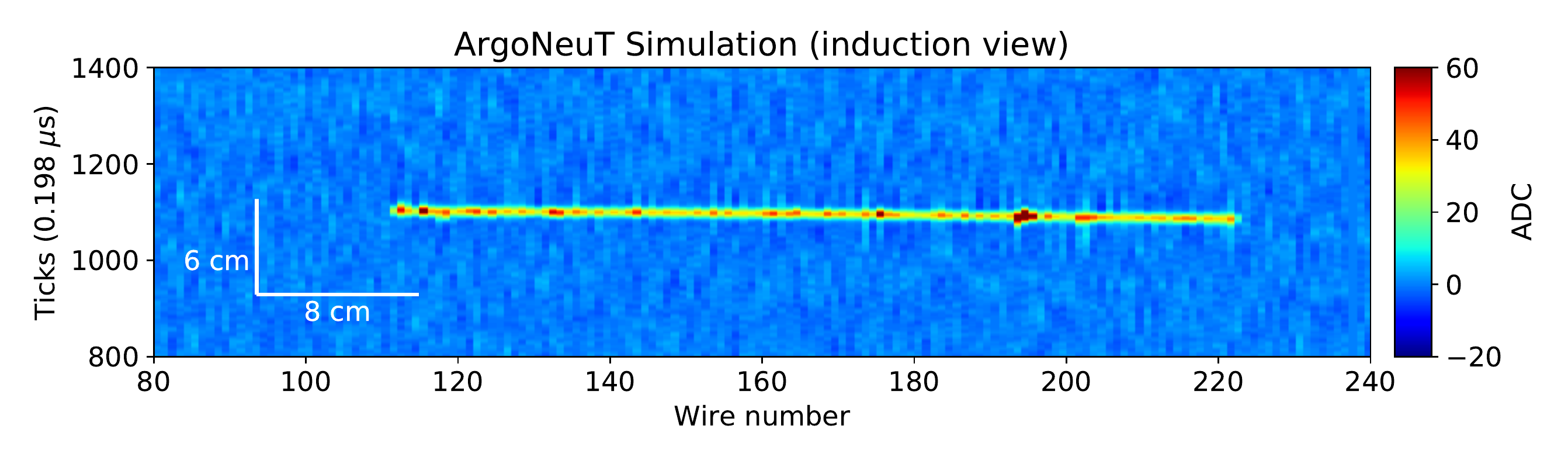}
  \includegraphics[width=.5\textwidth,keepaspectratio]{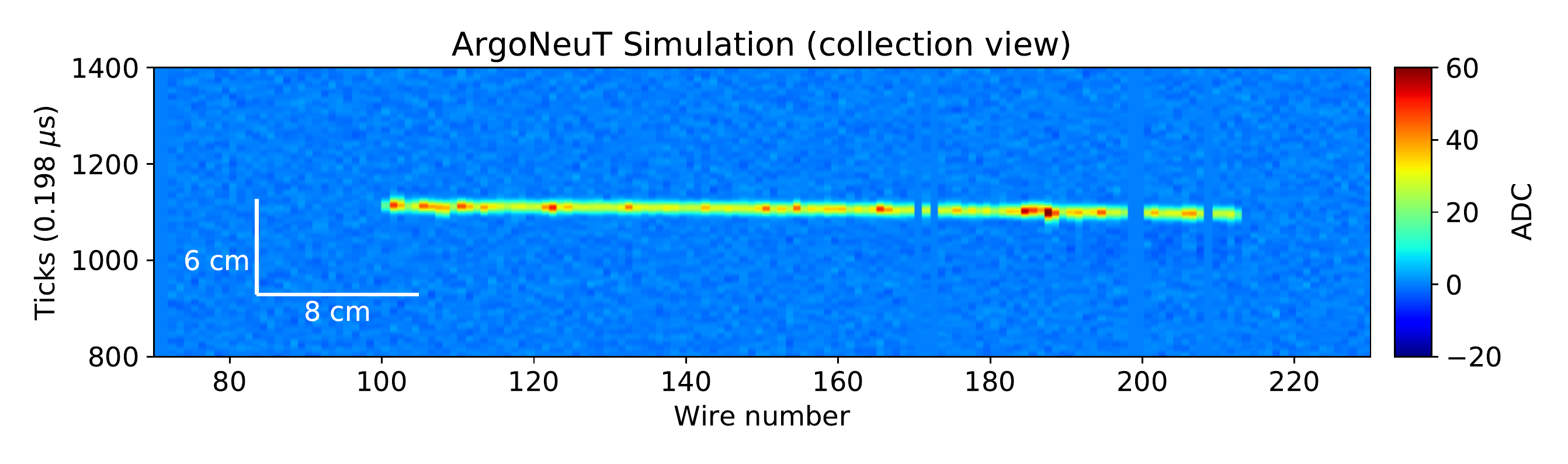}
  \includegraphics[width=.45\textwidth,keepaspectratio]{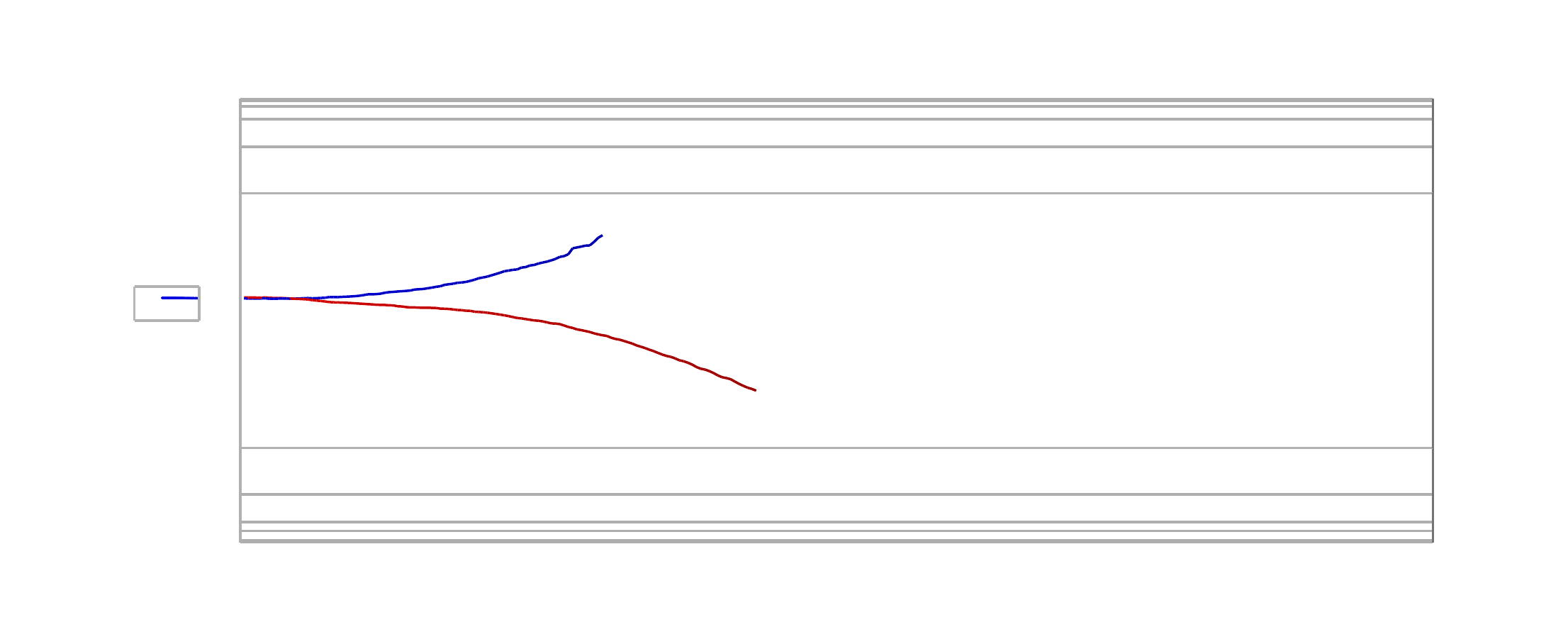}
  \caption{Event display of a simulated HNL decay with a double-MIP signature. The induction (top) and collection (middle) wire-plane views are shown in ArgoNeuT, where a single track is reconstructed. The color is proportional to the charge deposited. The single track is matched to a pair of tracks in the MINOS-ND (bottom), that are reconstructed with opposite charges represented by the different colors.}
  \label{fig:doubleMIPEvent}
\end{figure}

\begin{figure}%[h]
  \centering
  \includegraphics[width=.45\textwidth,keepaspectratio]{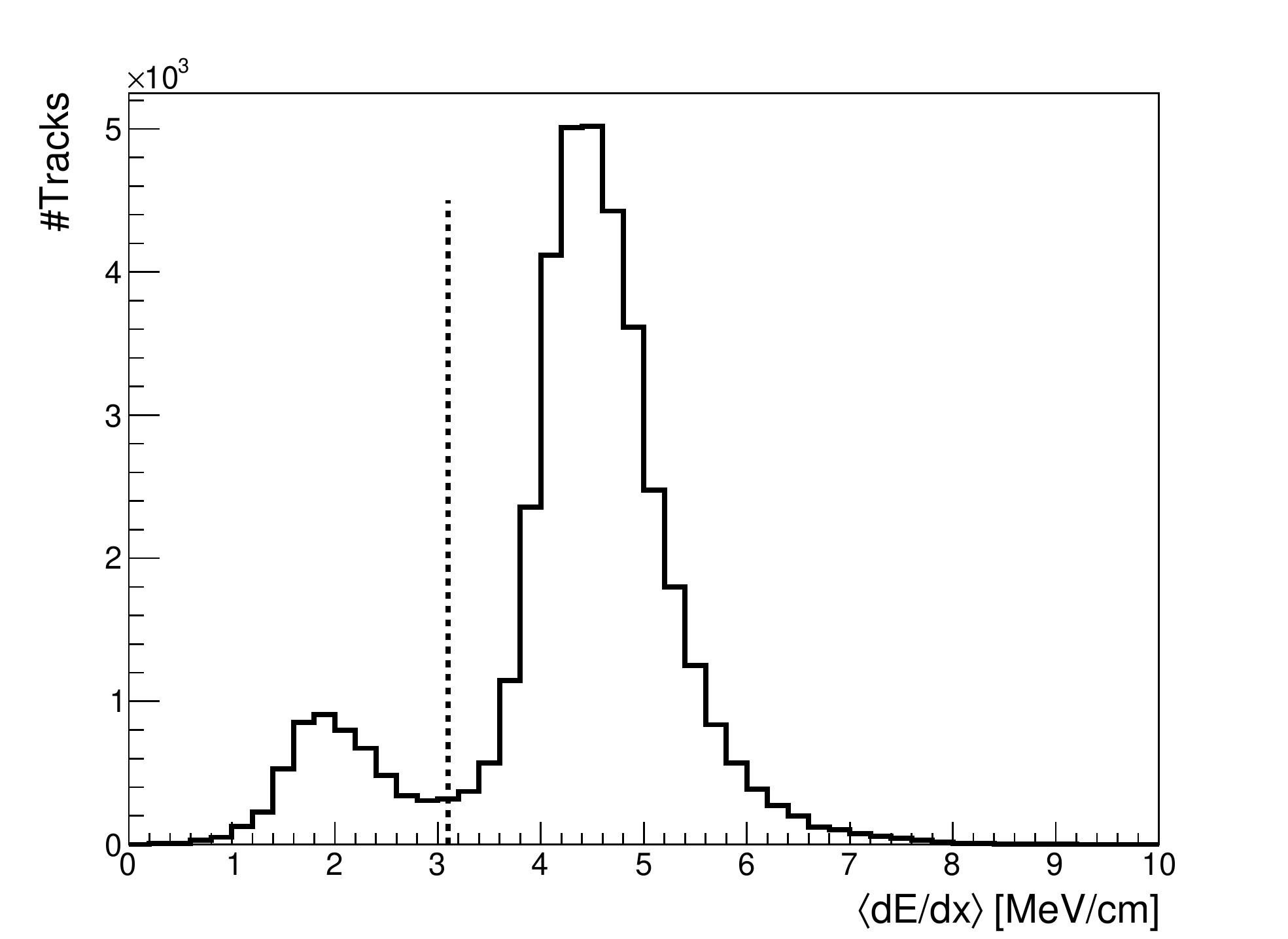}
  \caption{Average reconstructed $dE/dx$ over the first $5$\,cm of tracks resulting from simulated HNL decays. Two peaks are visible, one corresponding to single muons and the other corresponding to two overlapping muons. The threshold applied at $3.1$\,MeV/cm is illustrated by a dashed line.}
  \label{fig:dEdx}
\end{figure}

An example simulated HNL decay with a double-MIP signature is shown in Fig.~\ref{fig:doubleMIPEvent}. The pair of muon tracks fully overlap in ArgoNeuT (top, middle), then split once reaching the MINOS-ND (bottom) due to the magnetic field. The strongest identifier of whether a pair of overlapping muons are present is provided by the $dE/dx$ of the track. The region of interest is the start of each track, prior to the muon pair possibly splitting. Fig.~\ref{fig:dEdx} shows the average reconstructed $dE/dx$ over the first $5$\,cm of tracks resulting from simulated HNL decays. Two distinct peaks are visible. The first is at $dE/dx \sim 2$\,MeV/cm, approximately the $dE/dx$ of a single minimally-ionizing muon. For these events the opening angle of the muons is sufficiently large to properly reconstruct them as two separate tracks. The second peak is at $dE/dx \sim 4.5$\,MeV/cm, approximately double the single MIP $dE/dx$, indicating two overlapping muons. A threshold is applied between the two peaks, illustrated by the dashed line, separating the double-MIP-like and MIP-like populations. 

\emph{Selection}.--- 
A series of pre-selection cuts are first applied to remove poorly reconstructed events along with obvious non-HNL interactions. The highly forward-going muons from HNL decays can be challenging to reconstruct correctly in LArTPC detectors. This is because the ionisation tracks are near parallel to the readout planes and hence the drifted ionisation charge arrives on the wires at approximately the same time. These events may have large regions missed during the reconstruction and cannot be reliably identified. Therefore, we first remove events with fewer than 80\% of reconstructed energy depositions associated with reconstructed tracks. Next, events with more than three total tracks with length $L\geq 5$\,cm, or more than two tracks originating from a vertex are removed. This removes any events that are obvious non-HNL interactions due to having additional reconstructed particles present. A harsher cut requiring only two tracks to be present is not applied because a common failure mode of the reconstruction is the presence of split tracks in the region where the overlapping muons begin to separate. Tracks shorter than $L = 5$\,cm are not considered to avoid removing events that have short $\delta$-rays originating from the muons. Events passing the pre-selection are then assessed against the two-track and double-MIP selection criteria sequentially.

In the \textit{two-track} scenario only tracks starting within a fiducial volume in ArgoNeuT are considered, defined as: $1 \leq x \leq 46$\,cm (drift), $-19 \leq y \leq 19$\,cm (vertical) and $z \geq 3$\,cm (beam direction), to remove backgrounds originating from the cavern. Events with two tracks that either originate from or can be projected back to a common vertex within the fiducial volume are selected. The tracks are required to be forward-going with respect to the beam direction, have length $L \geq 5$\,cm, exit ArgoNeuT towards the MINOS-ND, have a mean $dE/dx$ over their full length consistent with being a single MIP ($dE/dx < 3.1$\,MeV/cm) and have an opening angle between them of $\theta_{opening} \leq 10 \degree$.

In the \textit{double-MIP} scenario, HNL decays occurring both inside ArgoNeuT and in the upstream cavern are considered. Any events containing tracks with an angle with respect to the beam direction $\theta_{beam} > 15$ are removed, as these likely originate from background interactions. The average $dE/dx$ is then calculated over the first 10 hits ($\sim 5$\,cm) of each track, where any individual anomalously large hits ($dE/dx > 10$\,MeV/cm) are discarded. A cut is applied at $dE/dx > 3.1$\,MeV/cm, illustrated by the dashed line in Fig.~\ref{fig:dEdx}, to identify events with a possible pair of overlapping muons.

Once candidate events are identified with either the two-track or double-MIP signature in ArgoNeuT, MINOS-ND matching is performed. Each track is projected to the start of the MINOS-ND and the radial and angular off-sets between the projected tracks and each reconstructed MINOS-ND track are compared. In the two-track case, ArgoNeuT-MINOS-ND matching tolerances of $r_{diff} \leq 12.0$\,cm and $\theta_{diff} \leq 0.17$\,rad are used \cite{Acciarri:2018ahy}. In the double-MIP case, since a single track is being matched to two tracks in the MINOS-ND, the matching tolerances are loosened to $2.5$ times the two-track case. The matched tracks are required to be forward-going with respect to the beam direction, start within $20$\,cm of the up-stream face of the detector and within the calorimeter region, and be at least $1$\,m long. This helps to remove any tracks that are unlikely to have originated from ArgoNeuT.

Finally, several selection cuts are applied in the MINOS-ND. These cuts are the same for both the two-track and double-MIP scenarios. We require that the tracks have an average $dE/dx$ consistent with being a muon ($4 \leq dE/dx \leq 18$\,MeV/cm), are reconstructed with opposite charges, and have start times, $t_0$, consistent with having originated from the same interaction or decay: $|\Delta t_0| \leq 20$\,ns. Pairs of track with larger $\Delta t_0$ could not have originated from a single HNL decay and instead are likely neutrino-induced background muons.

\begin{figure}%[h]
  \centering
  \includegraphics[width=.45\textwidth,keepaspectratio]{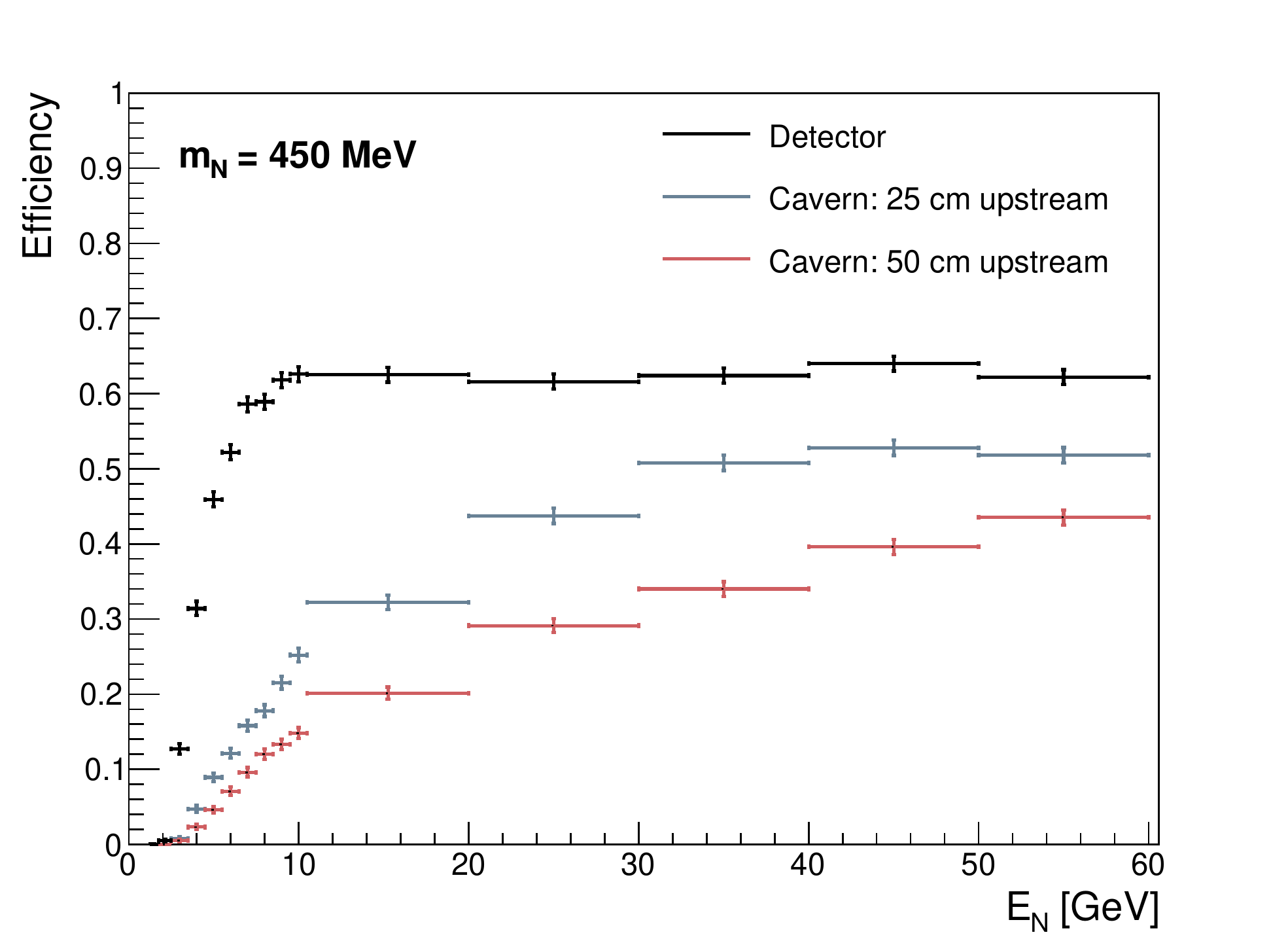}
  \caption{Selection efficiency as a function of $E_N$ for $m_N=450$\,MeV HNL decays occurring inside the ArgoNeuT detector (black) and at 25cm (blue) and 50cm (red) into the cavern upstream of ArgoNeuT along the beam direction.}
  \label{fig:efficiency}
\end{figure}

The selection efficiency as a function of the HNL energy, $E_N$, is shown in Fig.~\ref{fig:efficiency} for simulated $m_N=450$\,MeV HNL decays occurring inside the ArgoNeuT detector and at two positions in the upstream cavern. The efficiency inside the detector, defined as the fraction of events that are selected with either the two-track or double-MIP signatures, is around 60-65\% and relatively flat above $E_N \sim 10$\,GeV. However, it drops significantly at lower energies predominantly due to one or both of the muons being too low energy to reach the MINOS-ND. The cavern efficiencies are defined as the fraction of decays resulting in muons intersecting with the ArgoNeuT detector that are selected with the double-MIP signature. The further away from ArgoNeuT the decay occurs, the less likely the muon pair is to remain overlapping. This probability decreases further at lower energies where the muons are less forward-going.

\emph{Backgrounds and systematic uncertainties}.--- 
The primary backgrounds in this search originate from mis-reconstructed neutrino interactions occurring within the ArgoNeuT cryostat, and from neutrino-induced through-going muons arising from interactions upstream of the detector. Simulation of these backgrounds is performed with the GENIE~\cite{Andreopoulos:2009rq} neutrino event generator using NuMI beam fluxes provided by the MINERvA collaboration \cite{Aliaga:2016oaz}, along with a data-driven model of neutrino-induced through-going muons \cite{spitzthesis, ArgoNeuT:2011bms, Acciarri:2014isz}. In the two track scenario, the dominant form of observed background events are charged current $\nu_\mu$ interactions where either the interaction vertex or one or more tracks have been poorly reconstructed leading to these events not being removed. We expect to see $0.1 \pm 0.1$ events of this type in the data. In the double-MIP scenario, the dominant type of background events are single reconstructed muons that have either low energy $\delta$-rays or low energy protons near the track vertex causing them to have a double-MIP-like $dE/dx$. These can then be incorrectly matched to a pair of muons in the MINOS-ND if a second background muon is passing near the ArgoNeuT detector at approximately the same time. We expect to see $0.3 \pm 0.2$ events of this type in the data. The total expected background is therefore $0.4 \pm 0.2$ events.

The systematic uncertainties affecting the measurement are summarised in Table \ref{tab:systematics}. They are dominated by the uncertainty on the HNL flux. There is a 20\% uncertainty on the $D^\pm_{(s)}$ production \cite{Barlag:1988qj,Alves:1996rz,Lourenco:2006vw}. Then, the uncertainty on the branching ratios $D^\pm_{(s)} \rightarrow \tau^\pm + \nu_\tau$ \cite{Zyla:2020zbs} leads to an additional 5.7\% uncertainty on the $\tau^\pm$ flux. Combining these in quadrature leads to a 20.8\% uncertainty on the resulting HNL flux. 
Next, we consider the impact of uncertainties in the reconstruction by repeating the analysis with each parameter varied individually according to its assigned uncertainty. We apply uncertainties of 3\% on the tuning of the calorimetry~\cite{Acciarri:2018ahy}, 3\% on the track angular reconstruction~\cite{spitzthesis} and 6\% on the energy reconstruction of stopping particles in the MINOS-ND~\cite{Adamson:2009ju}. Finally, we assign a 1\% uncertainty on the charge reconstruction due to the modelling of the magnetic field~\cite{Adamson:2009ju}. Combining the impact of the performed variations in quadrature leads to a $0.5\%$ systematic uncertainty due to reconstruction effects. In addition to the reconstruction uncertainties, a 3.3\% systematic uncertainty is assigned to the selection efficiency to account for the potential impact of neutrino-induced through-going muons present in 3.3\% of triggers \cite{Anderson:2012vc,Anderson:2012mra}. A through-going muon registered in coincidence with a HNL event would lead to it being discarded in the pre-selection. Finally, there is a 2.2\% uncertainty in the size of the ArgoNeuT instrumented volume originating from uncertainty in the electron drift velocity \cite{spitzthesis} and a 1\% uncertainty in the number of collected POT \cite{Acciarri:2014isz}.

\begin{table}%[h]
    \centering
    \begin{tabular}{|c|c|}
    \hline
       Systematic Uncertainty & Impact (\%) \\\hline
       HNL flux & 20.8 \\
       Reconstruction effects & 0.5\\
       Selection efficiency & 3.3 \\
       Instrumented volume & $2.2$ \\
       POT counting & $1.0$\\ \hline
       Total & 21.2 \\ \hline
    \end{tabular}
    \caption{Systematic uncertainty impact on the sensitivity.}
    \label{tab:systematics}
\end{table}

\begin{figure}%[h]
  \centering
  \includegraphics[width=.45\textwidth,keepaspectratio]{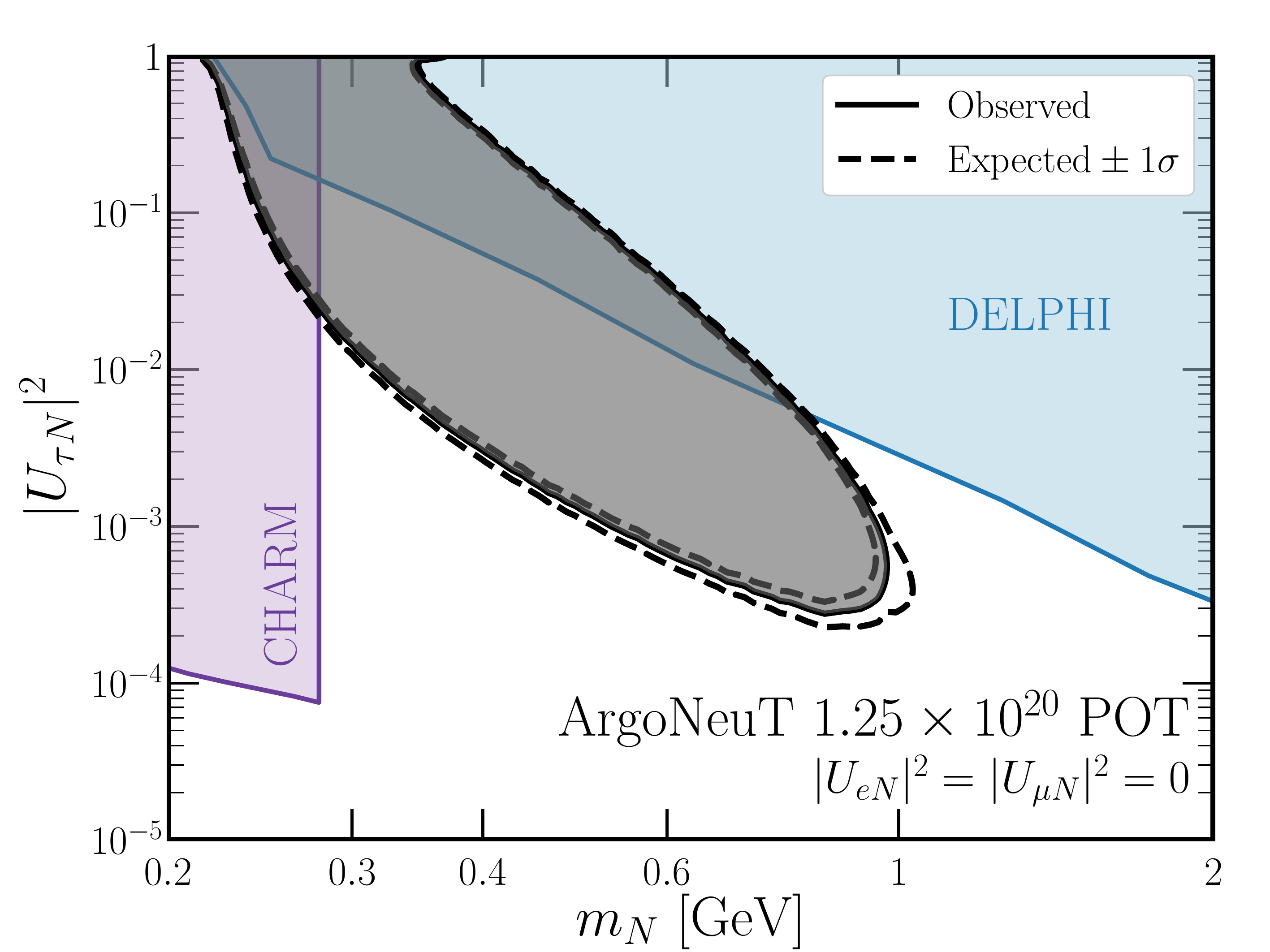}
  \caption{Constraints on the parameter space at 90\% CL from  $1.25 \times 10^{20}$ POT at ArgoNeuT (solid black), assuming production from $\Ptaupm$ decays. The dashed black lines show the uncertainty on the expected constraint at $\pm 1 \sigma$. The existing limits from CHARM \cite{Orloff:2002de} and DELPHI \cite{Abreu:1996pa} are also shown in purple and blue, respectively.}
  \label{fig:sensitivity}
\end{figure}

\emph{Results}.---
The selection has been applied to the full ArgoNeuT \num{1.25e20} POT anti-neutrino mode data-set. In total zero events pass, consistent with the expected background rate of $0.4 \pm 0.2$ events. Figure \ref{fig:sensitivity} shows our exclusion of parameter space at 90\% confidence level with $1.25 \times 10^{20}$ POT at ArgoNeuT, assuming production from $\Ptaupm$ decays. The limit is evaluated using a Bayesian approach with a uniform prior~\cite{Helene:1982pb}. The $\pm 1\sigma$ uncertainty on the expected constraint includes both the uncertainty on the background expectation and the $21.2\%$ systematic uncertainty on the signal production, combined conservatively. The existing limits from CHARM \cite{Orloff:2002de} and DELPHI \cite{Abreu:1996pa} are also shown in purple and blue, respectively. Our result leads to a significant increase in the exclusion region on the mixing angle $\left\vert U_{\tau N}\right\rvert^2$ of tau-coupled Dirac HNLs with masses $m_N =$ 280 - 970\,MeV, assuming $\left\vert U_{eN}\right\rvert^2 = \left\vert U_{\mu N}\right\rvert^2 = 0$. Other scenarios are considered in the Supplemental Material~\cite{supp}.  

\emph{Conclusions}.--- 
We have presented the first search for HNLs decaying with the signature $N \to \nu \mu^+ \mu^-$ in a LArTPC detector. Applying a novel technique to identify pairs of overlapping highly forward-going muons, we have searched for tau-coupled Dirac HNLs produced in the NuMI beam and decaying in the ArgoNeuT detector or in the upstream cavern. In the data, corresponding to an exposure to $\num{1.25e20}$ POT, zero passing events are observed consistent with the expected background. The results of this search lead to a significant increase in the exclusion region on the mixing angle $\left\vert U_{\tau N}\right\rvert^2$ of tau-coupled Dirac HNLs with masses $m_N =$ 280 - 970\,MeV, assuming $\left\vert U_{eN}\right\rvert^2 = \left\vert U_{\mu N}\right\rvert^2 = 0$. The analysis techniques we developed could be applied in future HNL searches performed in larger mass LArTPC experiments.

\emph{Acknowledgements}.--- 
This manuscript has been authored by Fermi Research Alliance, LLC under Contract No. DE-AC02-07CH11359 with the U.S. Department of Energy, Office of Science, Office of High Energy Physics. We gratefully acknowledge the cooperation of the MINOS Collaboration in providing their data for use in this analysis. We wish to acknowledge the support of Fermilab, the Department of Energy, and the National Science Foundation in ArgoNeuT’s construction, operation, and data analysis. This project has received funding from the Science and Technology Facilities Council (STFC), part of the United Kingdom Research and Innovation; and from the Royal Society UK awards: RGF\textbackslash EA\textbackslash 180209 and UF140089. We thank Dr Leigh Whitehead for his assistance processing the MINOS-ND data.

\bibliography{main.bib}

\begin{thebibliography}{45}
\bibitem [{foo()}]{footnote:1}%
  \BibitemOpen
  \href@noop {} {\bibinfo {title} {{$K(m_N) < 1$ enforces that this vanishes as
  $m_N$ grows larger than ${\sim}$\,GeV. The factor of $0.9$ accounts for the
  ${\sim}10\%$ of $\tau^\pm$ decays into heavier final states that would not
  allow for the production of a HNL in our mass range of
  interest.}}}\BibitemShut {Stop}%
\end{thebibliography}

\end{document}

\title{Supplemental material}
%\title{Supplemental material to the manuscript ...}
\maketitle

\section{Alternative Model Scenarios}

Here we consider two alternative scenarios where the only non-zero mixing angle between the Heavy Neutral Lepton, $N$, and the light neutrinos, $\nu_\alpha$ ($\alpha = e,\mu,\tau$), is either $\left\lvert U_{eN}\right\rvert^2$ or $\left\lvert U_{\mu N}\right\rvert^2$. In each case we include two production channels, $K^\pm \to \ell_\alpha^\pm N$ and $D_{(s)}^\pm \to \ell_\alpha^\pm N$ and, as before, the production rate of $N$ is proportional to the mixing angle squared $\left\lvert U_{\alpha N}\right\rvert^2$. Details about these production mechanisms can be found in Ref.~\cite{Coloma:2020lgy}. The $K^\pm$ channel is simulated using a GEANT4~\cite{Agostinelli:2002hh} based simulation of the NuMI beam (G4NuMI), allowing for decay-in-flight of the focused $K^\pm$. In this case, $N$ production is considered from interactions in the target only. The promptly decaying $D_{(s)}^\pm$ channel is simulated using PYTHIA8~\cite{Sjostrand:2014zea} as in the main text, considering production in both the target and the absorber. The signal considered in these scenarios is the same as in the main text, $N \to \nu \mu^+ \mu^-$. However, the accepted HNL flux from these production mechanisms is different due to the different energy distribution of the parent particles that the HNLs originate from. This is accounted for using the energy-dependent efficiencies for the HNL selection discussed in the main text with the different energy distributions.

The systematic uncertainties on the HNL flux in the electron- and muon-coupled scenarios are different to the tau-coupled scenario discussed in the main text. In the case of HNLs originating from $K^\pm$ decays, the uncertainty on the $K^\pm$ production is $^{+1.9}_{-5.8}\%$~\cite{Aduszkiewicz:2019xna}. In the case of HNLs originating from $D_{(s)}^\pm$ decays, the uncertainty on the $D_{(s)}^\pm$ production remains the same, but the uncertainty on the resulting HNL flux is no longer impacted by the uncertainty on the $D^\pm_{(s)} \rightarrow \tau^\pm + \nu_\tau$ branching ratio. The remaining systematic uncertainties are the same as discussed in the main text.

\begin{figure*}[h]
\centering
  \includegraphics[width=.45\textwidth,keepaspectratio]{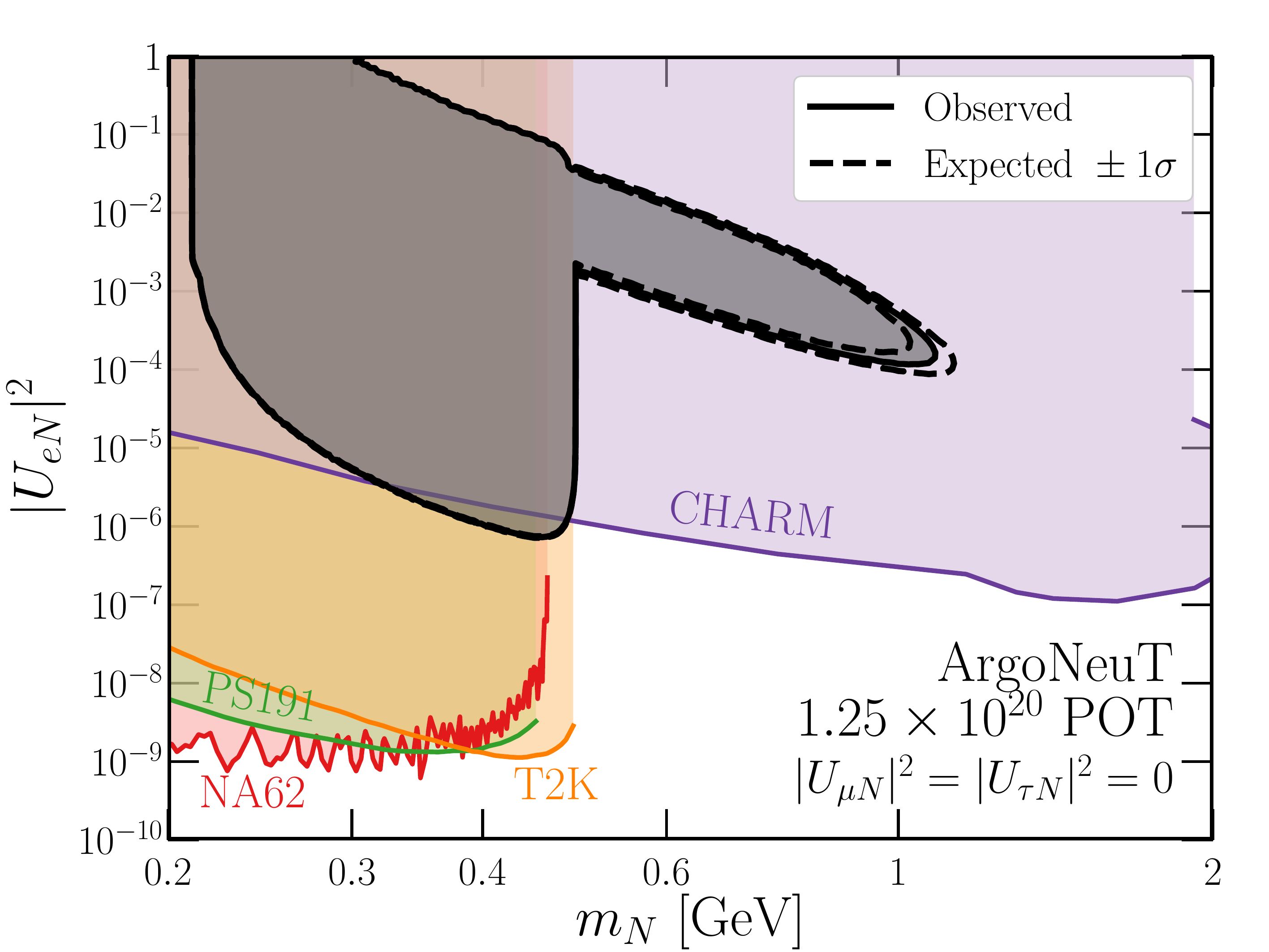}
  \includegraphics[width=.45\textwidth,keepaspectratio]{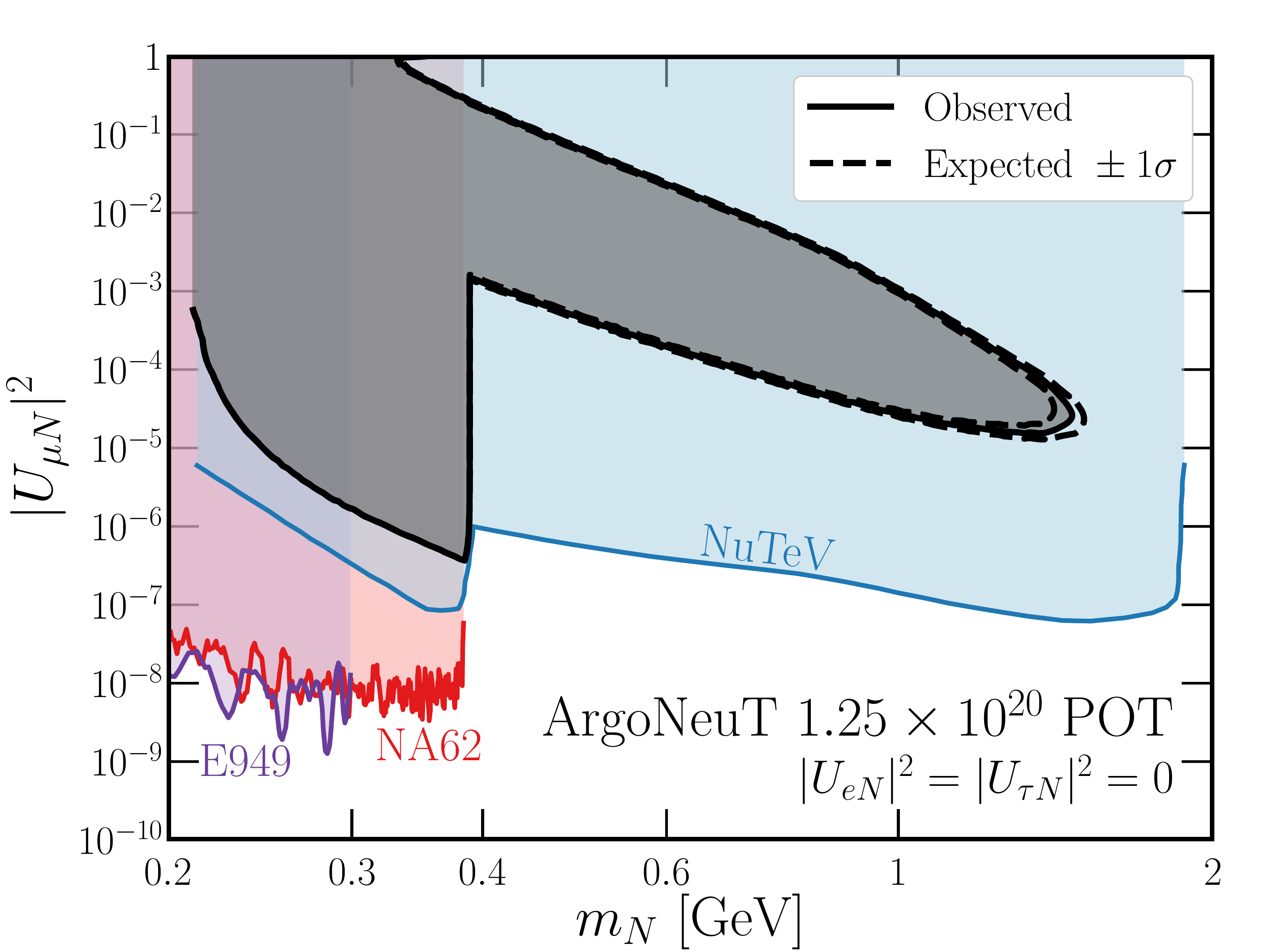}
  \caption{Constraints on the parameter space at 90\% CL from  $1.25 \times 10^{20}$ POT at ArgoNeuT (solid black) as a function of the HNL mass and mixing angle $\left\lvert U_{e N}\right\rvert^2$ (left) or $\left\lvert U_{\mu N}\right\rvert^2$ (right), assuming the other mixing angles are zero in each case. The dashed black lines show the uncertainties on the expected constraints at $\pm 1 \sigma$. The existing limits in each case are shown, see text for more detail.}
\label{fig:sensitivity:NeNmu}
\end{figure*}

Figure \ref{fig:sensitivity:NeNmu} shows our exclusion of parameter space at 90\% confidence level in the scenarios where $\left\lvert U_{eN}\right\rvert^2$ (left) and $\left\lvert U_{\mu N}\right\rvert^2$ (right) are the only non-zero mixing angles. The uncertainties on the expected constraints at $\pm 1 \sigma$ assuming zero observed events are shown by the dashed black lines. In each case the strongest existing constraints are shown. In the electron-coupled scenario, these are from NA62 (red)~\cite{NA62:2020mcv}, PS191 (green)~\cite{Bernardi:1985ny,Bernardi:1987ek}, T2K (orange)~\cite{Abe:2019kgx}, and CHARM (purple)~\cite{Bergsma:1985is}. In the muon-coupled scenario, these are from E949 (purple)~\cite{Artamonov:2009sz}, NA62 (red)~\cite{CortinaGil:2021gga}, and NuTeV (blue)~\cite{Vaitaitis:1999wq}. In both cases, less sensitive constraints are not shown to improve clarity. We find that the extracted limits from ArgoNeuT are considerably weaker than other existing limits. This is due to two factors: greater exposure and access to other final-states with larger branching ratios, such as $N \to e^- \pi^+$ in the electron-coupled case.   

\bibliography{supp.bib}